\documentclass[twoside,12pt]{article}  %%% ���� LaTeX ��ʽ�ļ�������ָ�����ȱ��
\usepackage{CJK}   %%%%% ʹ�����ġ����ġ�����ʱ���ô˺���
\usepackage{amsmath}
\usepackage{indentfirst}  %% �ڱ�������һ�ж�����
\usepackage{bm}    %%%%%  ��ӡ��ѧ���ź�б��ʱ���ô˺���
\usepackage{graphicx}  %%%% ���в��� eps ͼ��ʱ���ô˺�������ͼ���� 1
\usepackage[usenames,dvipsnames]{color} %%% ��ɫ����
\begin{document}    %% �ı��ļ���ʼ�����Ǳ�����ָ��

\begin{CJK*}{GBK}{song}  %% ��ʼ�������Ļ���

%=================== Text begin here =============================================

\begin{center}
\LARGE\bf Confined subdiffusion in three dimensions{\color{blue}$^{*}$}   %% ������Ŀ
\end{center}

\footnotetext{\hspace*{-.45cm}\footnotesize $^*$Project supported by National
Natural Science Foundation of China (Grant No. 21153002), and the Fundamental Research Funds for the Central Universities of Central South University (Grant No. 2013zzts151).}

\footnotetext{\hspace*{-.45cm}\footnotesize $^\dag$Corresponding author. E-mail: {\color{blue} yonghe@csu.edu.cn}}

\begin{center}
\rm Shanlin Qin and  \ Yong He {\color{blue}$^{\dagger}$}
\end{center}

\begin{center}
\begin{footnotesize} \sl
 Hunan Key Laboratory for Super-microstructure and Ultrafast Process,
 School of Physics and Electronics, Central South University,
 Changsha 410012, Hunan, China
%${}^{\rm b)}$ \\   %%%% ��ַ b)
%${}^{\rm c)}$ \\   %%%% ��ַ c)
%%% ������ַ�����������
\end{footnotesize}
\end{center}

\begin{center}
\footnotesize (Received X XX XXXX; revised manuscript received X XX XXXX)
          %% (Received �� �� ��; revised manuscript received �� �� ��)
\end{center}

\vspace*{2mm}

\begin{center}
\begin{minipage}{15.5cm}
\parindent 20pt\footnotesize
The three-dimensional (3D) Fick's diffusion equation and fractional diffusion equation are solved for different reflecting boundaries. We use the continuous time random walk model (CTRW) to investigate the time averaged mean square displacement ( MSD ) of 3D single particle trajectory. Theoretical results show the ensemble average of the time averaged MSD can be expressed analytically by a Mittag-Leffler function. Our new expression is in agreement with previous formulas in two limiting cases which are $ < \overline{\delta^2} > \sim\Delta $ in short lag time and $ < \overline{\delta^2}> \sim\Delta^{1-\alpha}$ in long lag time. We also simulate the experimental data of mRNA diffusion in living E. coli using 3D CTRW model under confined and crowded conditions. The simulated results are well consistent with experimental results. The calculations of power spectral density (PSD) indicate further the subdiffsive behavior of individual trajectory.
\end{minipage}
\end{center}

\begin{center}
\begin{minipage}{15.5cm}
\begin{minipage}[t]{2.3cm}{\bf Keywords:}\end{minipage}
\begin{minipage}[t]{13.1cm}
confined subdiffusion, three dimensions, time averaged mean squared displacement
\end{minipage}\par\vglue8pt
{\bf PACS: }
%%% PACS ������
%% ��ѯ��ַ��http://www.aip.org/pacs
\end{minipage}
\end{center}

\section{Introduction}  %%% �ڱ��� 1
Inside cells, the motion of the real biomolecules occurs in three dimensions. The three-dimensional (3D) trajectory tracking bring valuable information that is losing by two-dimensional (2D) tracking. Therefore, for an accurate determination of the diffusion, a 3D trajectory and analysis is required. Current the techniques of three-dimensional (3D) single particle tracking  ( SPT ) $^{[{\color{blue} 1-10}]}$ enable us to observe the motion of single particle with the position resolution of 3 nm and the time resolution of 100 ms $^{[{\color{blue} 4}]}$. But so far, the interpretation of experimental data was mostly restricted to one dimension. It is clear that there is a need for theory and simulation of microscopic models that can make quantitative predictions of the diffusion behaviour in three dimensions.

The diffusion environment of the single biomolecules in living cell is confined and crowded. Roughly estimation, the sizes of the single biomolecules and the cell are from a few to several hundred nanometers and from one to one hundred micrometers, respectively. During diffusion, when a jumping biomolecule meets cellular inner membrane, it will be reflected. So the diffusion is a confined motion. The cellular interior is also highly crowded. For example, a typical E. coli cell, its geometrical size is about 1um $\times$ 1um $\times$ 2um, the volume V $\approx$ 1um$^3$. Inside E. coli cell, there are about $2 \times 10^6$ proteins, $ 2 \times 10^4$ ribosomes and $ 2 \times 10^{10}$ water molecules. The mean spacing between protein molecules within E. coli cell is less than 10 nanometers $^{[{\color{blue} 11}]}$.

The motion of single biomolecules inside cell often exhibits subdiffusion with slow diffusion coefficient in  the confined and crowded  environment. Up to now, for the interpretation of experimental results, three theoretical models are commonly employed $^{[{\color{blue} 12-16}]}$. The first approach is Gaussian models like fractional Brownian motion $^{[{\color{blue} 17-20}]}$ and Langevin equations $^{[{\color{blue} 21-24}]}$, the second category is the continuous-time random walk $^{[{\color{blue} 25-28}]}$, and the last method is obstructed diffusion $^{[{\color{blue} 29-32}]}$. However, a thorough understanding of microscopic mechanism of the single biomolecules diffusion is still a challenge $^{[ {\color{blue} 12-16,33-34}]}$.

We have applied the model of continuous time random walk (CTRW) simulated $^{[{\color{blue} 35-36}]}$ the experimental results $^{[{\color{blue} 37}]}$ on the diffusion of mRNA molecules inside live E. coli.
Recently, the large numbers of theoretical works have made remarkable headway in investigations on subdiffusion, especially on ageing and weak ergodicity breaking$^{[{\color{blue} 14-15, 38-45}]}$. Here we extend previous 1D model $^{[{\color{blue} 35,39}]}$ to 3D. We pay especially attention to that case with different boundary conditions in three spatial directions.

The time-average mean squared displacement (MSD) $\overline{\delta_l^2}$ of the lth trajectory is defined through three-dimension trajectory $\overrightarrow{r}(t)$, which is recorded in the time interval $(0, T)$, according to
\begin{equation}
\begin{array}{l l l l }
\overline{\delta_l^2 } \left(\Delta, T \right) =
{1 \over T - \Delta} {\int_{0} ^{ T - \Delta} \left[ \overrightarrow{r}(t + \Delta) - \overrightarrow{r} ( t ) \right]^2 {\rm d } t },
\label{eq01}
\end{array}
\end{equation}
where l = 1,2,3 ... N labels the number of every different trajectory, in which T is the finite measurement time, $ \Delta $ denotes the lag time. The $\overline{\delta_l^2}$ can be obtained from a trajectory. From all N trajectories, we can get the ensemble average of the time averaged MSD
\begin{equation}
\begin{array}{l l l l}
\langle \overline{\delta^2 } \left(\Delta, T \right) \rangle & = & {1 \over N} {\displaystyle\sum_{l=1}^{N}} \overline{\delta_l^2 } \left(\Delta, T \right) & \\
& = &  {1 \over T - \Delta} {\int_{0} ^{ T - \Delta} \langle \left[ \overrightarrow{r}(t + \Delta) - \overrightarrow{r} ( t ) \right]^2 \rangle {\rm d } t }, &
\end{array}
\label{eq02}
\end{equation}
above angular brackets $\langle \cdot \rangle$ denotes the ensemble average. The ensemble averaged of a random walk can also be calculated in another way,
\begin{equation}
\begin{array}{l l l l}
\langle \left[ \overrightarrow{r}(t + \Delta) - \overrightarrow{r} ( t ) \right]^2 \rangle & = &
\int_V\left[ \overrightarrow{r}(t + \Delta) - \overrightarrow{r} ( t ) \right]^2 P(\overrightarrow{r};t)
{\rm d }\overrightarrow{r},
\end{array}
\label{eq03}
\end{equation}
where the probability density function (PDF) $P(\overrightarrow{r};t)$ is the relative probability of finding the walker the at position $\overrightarrow{r}$ at time $t$. The $P(\overrightarrow{r};t)$ obeys the fractional diffusion equation$^{[{\color{blue} 26}]}$ in three dimensions,
\begin{equation}
\begin{array}{l l l l}
{\partial \over \partial t} P(\overrightarrow{r};t)
& = & _0D_t^{1-\alpha} K_\alpha \bigtriangledown^2 P(\overrightarrow{r};t),
\end{array}
\label{eq04}
\end{equation}
where $K_\alpha$ is the diffusion constant, and the Riemann-Liouville operator is defined by $^{[{\color{blue} 46}]}$
\begin{equation}
\begin{array}{l l l l}
 _0D_t^{1-\alpha} f(t) & = & {1 \over \Gamma ( \alpha ) } { \partial \over \partial t}
 \int_{0}^{t} {f(t^{'}) \over (t-t^{'})^{1-\alpha} } {\rm d } t^{'}.
\end{array}
\label{eq05}
\end{equation}

We suppose that the cell has a cuboid shape with three side lengths $L_x, L_y$ and $ L_z $ in the rectangular coordinate system. The boundary condition and initial condition are imposed as
\begin{equation}
\left\{\begin{array}{l l l l}
P(\overrightarrow{r}; t)|_{t=t_s} & = & \delta(x-x_s)\delta(y-y_s)\delta(z-z_s)), \\
{\partial \over \partial x} P(\overrightarrow{r}; t) |_{x=0, x=L_x} & = & 0, \\
{\partial \over \partial y} P(\overrightarrow{r}; t) |_{y=0, y=L_y} & = & 0, \\
{\partial \over \partial z} P(\overrightarrow{r}; t) |_{z=0, z=L_z} & = & 0.
\end{array} \right. \\
\label{eq06}
\end{equation}

In this paper, we solve the three-dimensional (3D) Fick's diffusion equation and fractional diffusion equation with different reflecting boundaries. We use the continuous time random walk model (CTRW) to explore the time averaged MSD of 3D single particle trajectory. Theoretical results show the ensemble average of the time averaged MSD can be expressed analytically by a Mittag-Leffler function. Our new expression is in agreement with previous formulas in two limiting cases which are $ < \overline{\delta^2} > \sim\Delta $ in short lag time and $ < \overline{\delta^2}> \sim\Delta^{1-\alpha}$ in long lag time. We also simulate the experimental data of mRNA diffusion in living E. coli using 3D CTRW model under confined and crowded conditions. The simulated results are well consistent with experimental results.
 The calculations of power spectral density (PSD) indicate further the subdiffsive behavior of individual trajectory.

\section{Theoretical analysis}  %%% �ڱ��� 2

\subsection{The exact solution of 3D Fick's diffusion equation}  %%% �ӽڱ��⣬�� 2.1.
We start from solving the 3D Fick's diffusion equation to calculate the ensemble averaged MSD in three dimension, which will be used to obtain the ensemble average of time averaged MSD later. The probability of finding the walker at position $\overrightarrow{r}$ at time $t$, if the walker was at position $\overrightarrow{r_s}$ at time $t_s$, obeys Fick's diffusion equation:
\begin{equation}\label{eq07}
\begin{array}{l l l l}
\frac{\partial }{\partial t} P(x,y,x,x_s,y_s,z_s;t,t_s)
& = & D\bigtriangledown^2 P(x,y,x,x_s,y_s,z_s;t,t_s),
\end{array}
\end{equation}
where $D$ is the diffusion constant. According to the Einstein relation, $D$ can be expressed by $D= \langle \delta r^2 \rangle /2 \langle t_w \rangle$, where $ \langle \delta r^2 \rangle$ is the variance of the jump lengths and $ \langle t_w \rangle $ is the average waiting time for the normal diffusion. This function is reasonably holding when $r^2 \gg \langle \delta r^2 \rangle $ and $t\gg \langle t_w \rangle $.

When taken $t_{s}=0$ and $0<x<L_x$, $0<y<L_y$, $0<z<L_z$, the initial condition and boundary condition are imposed as Eq. ( {\color{blue} \ref{eq06}} ). Therefore, the solution of Eq. ( {\color{blue} \ref{eq07}} ) is derived as
\begin{equation}\label{eq09}
\begin{array}{l l l l}
P(x,y,z,x_s,y_s,z_s;t)  \\
=\sum_{n=0}^{\infty}\sum_{m=0}^{\infty}\sum_{k=0}^{\infty}\frac{2^{\delta_{m0}+\delta_{n0}+\delta_{k0}}}{L_xL_yL_z}cos(\frac{n\pi x}{L_x})cos(\frac{m\pi y}{L_y})cos(\frac{k\pi z}{L_z})\\
\times cos(\frac{n\pi x_s}{L_x})cos(\frac{m\pi y_s}{L_y})cos(\frac{k\pi z_s}{L_z})exp(-[(\frac{n\pi}{L_x})^2+(\frac{m\pi}{L_y})^2+(\frac{k\pi}{L_z})^2]Dt),
\end{array}
\end{equation}
where $\delta_{m0} (\delta_{n0}, \delta_{k0} )$ is called Kronecker's delta and satisfies the properties that when $m (n, k)=0$, $\delta_{m0} (\delta_{n0}, \delta_{k0} ) =1$ and $m (n, k) \neq0$, $\delta_{m0} (\delta_{n0}, \delta_{k0} ) =0$.

\subsection{The ensemble averaged MSD for normal diffusion}  %%% �ӽڱ��⣬�� 2.1.

For $r^2=x^2+y^2+z^2$, the ensemble averaged MSD of normal diffusion, $\langle r^2(t) \rangle$, can be derived from Eq. ( {\color{blue} \ref{eq09}} ) as
\begin{equation}\label{eq10}
\begin{array}{l l l l}
\langle r^2(t) \rangle = \frac{1}{L_xL_yL_z}\int_{0}^{L_x}\int_{0}^{L_y}\int_{0}^{L_z}{\rm d}x_s {\rm d}y_s{\rm d}z_s\int_{0}^{L_x}\int_{0}^{L_y}\int_{0}^{L_z}{\rm d}x{\rm d}y{\rm d}z\\
\times P(x,y,z,x_s,y_s,z_s;t)[(x-x_s)^2+(y-y_s)^2+(z-z_s)^2]\\
=\frac{L_x^2+L_y^2+L_z^2}{6}+\sum_{n=0}^{\infty}\frac{16L_x^2}{(2n+1)^4\pi^4}exp[-\frac{(2n+1)^2\pi^2Dt}{L_x^2}]\\
+\sum_{m=0}^{\infty}\frac{16L_y^2}{(2m+1)^4\pi^4}exp[-\frac{(2m+1)^2\pi^2Dt}{L_y^2}]\ \\
+\sum_{k=0}^{\infty}\frac{16L_z^2}{(2k+1)^4\pi^4}exp[-\frac{(2k+1)^2\pi^2Dt}{L_z^2}],\\
\end{array}
\end{equation}
when $t$ is large, the ensemble averaged MSD for normal diffusion reaches a constant $(L_x^2+L_y^2+L_z^2)/6$. The MSD of confined random walker after $j=t/\langle t_w \rangle$ jumps has the form
\begin{equation}\label{eqx2j}
\begin{array}{l l l l}
\langle r^2(j) \rangle = \sum_{i=0}^{\infty} \gamma_i q_i^j, \\
\end{array}
\end{equation}
with $q_0=1$ and $q_i < 1$. The coefficients $q_i$ and $\gamma_i$ can be derived by comparison with Eq. ( {\color{blue} \ref{eq10}} ).

\subsection{The exact solution of 3D fractional diffusion equation}  %%% �ӽڱ��⣬�� 2.1.
We can solve the 3D fractional diffusion equation by using the separation of variables $^{[{\color{blue} 39, 47}]}$. Let
\begin{equation}
\begin{array}{l l l l}
P(\overrightarrow{r}; t) & = &  X(x)Y(y)Z(x)T(t),
\end{array}
\label{eq11}
\end{equation}
and substitute back into Eq. ( {\color{blue} \ref{eq04}} ), we obtain
\begin{equation}
\left\{\begin{array}{l l l l}
{\partial^2 f( q ) \over \partial q^2} & = & \lambda_q f(q), \\
f (q) |_{t=t_s} & = & \delta(q - q_s), \\
{\partial \over \partial q} f(q) |_{q=0, q=L_q} & = & 0,
\end{array} \right.
\label{eq12}
\end{equation}
\begin{equation}
\left\{\begin{array}{l l l l}
{\partial \over \partial t} T(t) & = & - \lambda  _0D_t^{1-\alpha} T(t), \\
T (t) |_{t=t_s} & = & 1,
\end{array} \right.
\label{eq13}
\end{equation}
where the coordinate q = x, y, or z,  the  function f(q) = X(x), Y(y), or Z(z), and constant $\lambda = \lambda_x + \lambda_y + \lambda_z$.
The solution of the Eq. ( {\color{blue} \ref{eq04}} ) should be labeled according to the choice of constants n, m and k, that is
\begin{equation}
\begin{array}{l l l l}
P_{nmk}(\overrightarrow{r}; t) & = &  X_n(x)Y_m(y)Z_k(x)T_{nmk}(t).
\end{array}
\label{eq14}
\end{equation}
The general solution is a linear combination of solutions $ P_{nmk}$,
\begin{equation}
\begin{array}{l l l l}
P(\overrightarrow{r}; t) =  \displaystyle\sum\limits_{nmk=0}^{\infty}  C_{nmk} P_{nmk}(x,y,z; t)   \\
={1 \over L_xL_yL_z} \{1 + 2 [ {\displaystyle\sum_{n=1}^{\infty}}cos(\frac{n\pi x_s}{L_x})cos(\frac{n\pi x}{L_x})
 E_\alpha (-\frac{n^2\pi^2}{L_x^2}K_\alpha t^\alpha ) \\
+ {\displaystyle\sum_{m=1}^{\infty}}cos(\frac{m\pi y_s}{L_y})cos(\frac{m\pi y}{L_y})  E_\alpha (-\frac{m^2\pi^2}{L_y^2}K_\alpha t^\alpha ) \\
+ {\displaystyle\sum_{k=1}^{\infty}}cos(\frac{k\pi z_s}{L_z})cos(\frac{k\pi x}{L_x})  E_\alpha (-\frac{k^2\pi^2}{L_z^2}K_\alpha t^\alpha ) ] \\
+ 4 [ {\displaystyle\sum_{nm=1}^{\infty}}cos(\frac{n\pi x_s}{L_x})cos(\frac{n\pi x}{L_x})cos(\frac{m\pi y_s}{L_y})cos(\frac{m\pi y}{L_y}) \\
 \times E_\alpha [-(\frac{n^2\pi^2}{L_x^2}+\frac{m^2\pi^2}{L_y^2} ) K_\alpha t^\alpha ] \\
+ {\displaystyle\sum_{nk=1}^{\infty}}cos(\frac{n\pi x_s}{L_x})cos(\frac{n\pi x}{L_x})cos(\frac{k\pi z_s}{L_z})cos(\frac{k\pi z}{L_z}) \\
\times  E_\alpha [-(\frac{n^2\pi^2}{L_x^2}+\frac{k^2\pi^2}{L_z^2} ) K_\alpha t^\alpha ] \\
+ {\displaystyle\sum_{mk=1}^{\infty}}cos(\frac{m\pi y_s}{L_y})cos(\frac{m\pi y}{L_y})cos(\frac{k\pi z_s}{L_z})cos(\frac{k\pi z}{L_z}) \\
 \times E_\alpha [-(\frac{m^2\pi^2}{L_y^2}+\frac{k^2\pi^2}{L_z^2} ) K_\alpha t^\alpha ] ] \\
+ 8  {\displaystyle\sum_{nmk=1}^{\infty}}cos(\frac{n\pi x_s}{L_x})cos(\frac{n\pi x}{L_x})  \\
  \times cos(\frac{m\pi y_s}{L_y})cos(\frac{m\pi y}{L_y})cos(\frac{k\pi z_s}{L_z})cos(\frac{k\pi z}{L_z}) \\
  \times  E_\alpha [-(\frac{n^2\pi^2}{L_x^2}+\frac{m^2\pi^2}{L_y^2}+\frac{k^2\pi^2}{L_z^2} ) K_\alpha t^\alpha ] \},
\end{array}
\label{eq15}
\end{equation}
where $ E_\alpha $ is the Mittag-Leffler function$^{[{\color{blue} 26}]}$.

\subsection{The ensemble average MSD for fractional diffusion}
The initial positions $x_s$, $y_s$ and $z_s$ are also the stochastic variables. In the equilibrium state, the ensemble averaged MSD should be independent on the $x_s$, $y_s$ and $z_s$,
\begin{equation}
\begin{array}{l l l l}
\langle \left[ \overrightarrow{r}(t + \Delta) - \overrightarrow{r} ( t ) \right]^2 \rangle
 =  {1 \over L_xL_yL_z} \int_{0}^{L_x}\int_{0}^{L_y}\int_{0}^{L_z}{\rm d} x_s  {\rm d}y_s {\rm d} z_s\int_{0}^{L_x}\int_{0}^{L_y}\int_{0}^{L_z}  {\rm d }x {\rm d}y {\rm d}z \\
\times [(x-x_s)^2+(y-y_s)^2+(z-z_s)^2]  P(\overrightarrow{r}; t) \\
= {1 \over 6}(L_x^2+L_y^2+L_z^2) - {16 \over \pi^4} \{ L_x^2 { \displaystyle\sum_{k_x=0}^{\infty}} {1 \over (2k_x+1)^4} E_\alpha \{- {\pi^2(2k_x+1)^2 \over L_x^2} K_\alpha t^\alpha \}\\
+L_y^2 {\displaystyle\sum_{k_y=0}^{\infty}} {1 \over (2k_y+1)^4} E_\alpha \{- {\pi^2(2k_y+1)^2 \over L_y^2} K_\alpha t^\alpha \}\\
+L_z^2 { \displaystyle\sum_{k_z=0}^{\infty}} {1 \over (2k_z+1)^4} E_\alpha \{- {\pi^2(2k_z+1)^2 \over L_z^2} K_\alpha t^\alpha \} \}, \\
\end{array}
\label{eq16}
\end{equation}
when $t$ is large, the ensemble average MSD reaches a constant $(L_x^2+L_y^2+L_z^2)/6$.

\subsection{The exact solution of the ensemble average of time averaged MSD} %%% ���ӽڱ��⣬ �� 2.1.1.
Let $P_j(x,y,z, x_s,y_s,z_s)$, which is governed by the 3D Fick's diffusion equation, Eq. ( {\color{blue} \ref{eq07}} ), be the probability of walker from $x_s, y_s, z_s$ to $x , y, z$ after j jump events, and $\chi_j(\Delta,t_s)$ be the probability of walker make $j$ jump events in the time interval $\Delta$ starting from $t_s$. We can then express the probability of CTRW with two independent stochastic processes, which are for the displacements and the waiting times respectively, as$^{[{\color{blue} 39,48}]}$
\begin{equation}\label{eq17}
\begin{array}{l l l l}
P(x,y,z,x_s,y_s,z_s;t_s + \Delta,t_s) & = & \sum_{j=0}^{\infty}P_j(x,y,z, x_s,y_s,z_s)\chi_j(\Delta,t_s).
\end{array}
\end{equation}

The ensemble averaged MSD in $[t_s,t_s + \Delta]$ can be calculated using $P(x,y,z,x_s,y_s,z_s;t_s + \Delta,t_s)$. Therefore, by using Eq. ( {\color{blue} \ref{eq17}} ), one can get
\begin{equation}\label{eq18}
\begin{array}{l l l l}
\langle [\overrightarrow{r}(t_s + \Delta)-\overrightarrow{r}(t_s)]^2 \rangle \\
=\int_V\left[ \overrightarrow{r} (t_s + \Delta) - \overrightarrow{r}( t_s ) \right]^2 P(x,y,z,x_s,y_s,z_s;t_s + \Delta,t_s) {\rm d }\overrightarrow{r}\\
=\sum_{j=0}^{\infty} \langle r^2(j) \rangle \chi_j(\Delta,t_s).\\
\end{array}
\end{equation}
Noting that $\langle r^2(j) \rangle $ in Eq. ( {\color{blue} \ref{eq18}} ) can be found by substituting $P(x,y,z,x_s,y_s,z_s;t_s + \Delta,t_s)$ with $P_j(x,y,z,x_s,y_s,z_s)$ in Eq. ( {\color{blue} \ref{eq10}} ), we then find the ensemble average of time averaged MSD for $\Delta \le T$ as
\begin{equation}\label{eq19}
\begin{array}{l l l l}
\langle \overline{\delta^2}(\Delta,T) \rangle & = & \frac{1}{T-\Delta}\sum_{j=0}^{\infty} \langle r^2(j) \rangle \int_{0}^{T-\Delta}{\rm d}t_s\chi_j(\Delta,t_s).
\end{array}
\end{equation}

Let $w(t)$ be the PDF of waiting time, so in a certain time span $[t_s, t_s+\Delta]$, the probability of making $j$ jumps is the the product of two parts, making $j-1$ jumps in a shorter time interval $[t_s, t_s+t]$ and finding a waiting time $\Delta-t$, then integrated over all possible t. Specifically, the probability of
making $n\ge2$ jumps in the time span $[t_s, t_s+\Delta]$ is
\begin{equation}\label{eq20}
\begin{array}{l l l l}
\chi_j(\Delta,t_s) & = & \int_{0}^{\Delta}\chi_{j-1}(t,t_s)w(\Delta-t){\rm d}t.
\end{array}
\end{equation}
Likewise, the probabilities of making no or one jump can be given respectively as
\begin{equation}\label{eq21}
\begin{array}{l l l l}
\chi_1(\Delta,t_s) & = & \int_{0}^{\Delta}w_{1}(t,t_s)w(\Delta-t){\rm d}t,
\end{array}
\end{equation}
\begin{equation}\label{eq22}
\begin{array}{l l l l}
\chi_0(\Delta,t_s) & = & 1-\int_{0}^{\Delta}w_{1}(t,t_s){\rm d}t.
\end{array}
\end{equation}
After averaging over $t_s$ and by using the Laplace transformation $\Delta \rightarrow u$, one can get for $j\ge 1$ and $j=0$
\begin{equation}
\left\{\begin{array}{l l l l}
\overline{\chi_j}(u) & = & \frac{1-w(u)}{u}[w(u)]^{j-1}\overline{w_1}(u)], \\
\overline{\chi_0}(u) & = & [1-\overline{w_1}(u)]/u.
\end{array} \right.
\label{eq23}
\end{equation}

Now the ensemble average of time averaged MSD, Eq. ( {\color{blue} \ref{eq19}} ), can be expressed in the Laplace representation. Using the geometrical series, the Laplace representation of $ \langle \overline{\delta^2} \rangle $
within the confined CTRW is given by
\begin{equation}\label{eq24}
\begin{array}{l l l l}
\langle \overline{\delta^2}(u) \rangle  & = & \frac{\overline{w_1}(u)}{u}\frac{1-w(u)}{w(u)}\sum_{i=0}^{\infty}\frac{\gamma_i}{1-q_i w(u)}.
\end{array}
\end{equation}
For a PDF of waiting time distributed like $w(t)=\frac{\alpha}{(1+t)^{1+\alpha}}$, its Laplace transformation is
\begin{equation}\label{eq25}
w(u)\approx \alpha e^uu^{\alpha}(\Gamma(-\alpha)+\frac{e^{-u}u^{-\alpha}}{\alpha})\approx1-\Gamma(1-\alpha)u^\alpha.
\end{equation}
The initial waiting time distribution can be approximated as $\overline{w_1}(u)\approx(ut)^{\alpha-1}/\Gamma(1+\alpha)$, which is an approximation valid when $u\gg t^{-1}$.
Taking into account only the terms $n=m=k=0$, one can get
\begin{equation}\label{eq26}
\begin{array}{l l l l}
\langle r^2(t) \rangle = \frac{L_x^2+L_y^2+L_z^2}{6}+\frac{16L_x^2}{\pi^4}exp[-\frac{\pi^2Dt}{L_x^2}]\\
+\frac{16L_y^2}{\pi^4}exp[-\frac{\pi^2Dt}{L_y^2}]\
+\frac{16L_z^2}{\pi^4}exp[-\frac{\pi^2Dt}{L_z^2}],
\end{array}
\end{equation}
and compare it with
\begin{equation}\label{eq27}
\begin{array}{l l l l}
\langle r^2(j) \rangle & \approx &  \gamma_0q_0^j+\gamma_1q_1^j+\gamma_2q_2^j+\gamma_3q_3^j,
\end{array}
\end{equation}
one can get
\begin{equation}\label{eq28}
\left\{\begin{array}{l l l l }\gamma_0=(L_x^2+L_y^2+L_z^2)/6, \quad q_0=1,\\
\gamma_1=-L_x^2/6, \quad q_1=exp(-\frac{\pi^2<\delta r^2>}{2L_x^2}),\\
\gamma_2=-L_y^2/6, \quad q_2=exp(-\frac{\pi^2<\delta r^2>}{2L_y^2}),\\
\gamma_3=-L_z^2/6, \quad q_3=exp(-\frac{\pi^2<\delta r^2>}{2L_z^2}).
\end{array} \right.
\end{equation}
So we have
\begin{equation}\label{eq29}
\begin{array}{l l l l}
\langle \overline{\delta^2}(u) \rangle =\frac{u^{\alpha-2}}{6\Gamma(1+\alpha)T^{1-\alpha}}[L_x^2\frac{1-q_1}{1-q_1+q_1\Gamma(1-\alpha)u^\alpha}\\
+L_y^2\frac{1-q_2}{1-q_2+q_2\Gamma(1-\alpha)u^\alpha}+L_z^2\frac{1-q_3}{1-q_3+q_3\Gamma(1-\alpha)u^\alpha}].
\end{array}
\end{equation}
By the reason $\langle \delta r^2 \rangle \ll L_x^2 $ ($L_y^2$, $L_z^2$), we have $q_1\approx 1-\frac{\pi^2<\delta r^2>}{2L_x^2}$, $q_2\approx 1-\frac{\pi^2<\delta r^2>}{2L_y^2}$ and $q_3\approx 1-\frac{\pi^2<\delta r^2>}{2L_z^2}$. In Eq. ( {\color{blue} \ref{eq29}} ), we ignore more high order terms than $ u^\alpha $ after reduction of fractions to a common denominator, then obtain
\begin{equation}\label{eq30}
\begin{array}{l l l l}
\langle \overline{\delta^2}(u) \rangle & = & \frac{L_x^2+L_y^2+L_z^2}{6} \frac{u^{\alpha-2}}{\Gamma(1+\alpha)T^{1-\alpha}}\frac{1}{1+(\tau_c u)^\alpha},
\end{array}
\end{equation}
in which $\tau_c=(\frac{L_x^2+L_y^2+L_z^2}{\pi^2 K_\alpha})^{1/\alpha}$, $K_\alpha=\frac{\langle \delta r^2 \rangle }{2\Gamma(1-\alpha)}$.
\par
Taking the inverse Laplace transformation for Eq. ( {\color{blue} \ref{eq30}} ) and integration with the use of
$\int_{0}^{z}E_{\alpha,\beta}(\lambda t^\alpha)t^{\beta-1}{\rm d}t=z^{\beta}E_{\alpha,\beta+1}(\lambda z^\alpha),\beta>0$ $^{[{\color{blue} 49}]}$, we obtain the ensemble average of time averaged MSD as
\begin{equation}\label{eq31}
\begin{array}{l l l l}
\langle \overline{\delta^2}(\Delta) \rangle & = & \frac{L_x^2+L_y^2+L_z^2}{6}\frac{\Delta }{\Gamma(1+\alpha)\tau_{c}^{\alpha}T^{1-\alpha}}E_{\alpha,2}[-(\Delta/\tau_c)^{\alpha}].
\end{array}
\end{equation}

From Eq. ( {\color{blue} \ref{eq31}} ), we can evaluate in two limiting cases, corresponding to $\Delta \gg \tau_c$, the long $\Delta$ behavior, and $\Delta \ll \tau_c$, the short $\Delta$ behavior$^{[{\color{blue} 39,50,51}]}$. First, we start with the case $\Delta \gg \tau_c$. By utilizing the asymptotic behaviour of Mittag-Leffler function: $E_{\alpha,2}[-(\Delta/\tau_c)^\alpha]\sim\frac{1}{\Gamma(1-\alpha)(\Delta/\tau_c)^\alpha}$, when $\Delta \gg \tau_c$ and $0<\alpha<1$ $^{[{\color{blue} 52}]}$, we conclude that:
\begin{equation}\label{eq32}
\begin{array}{l l l l}
\langle \overline{\delta^2}(\Delta) \rangle & \approx & \frac{(L_x^2+L_y^2+L_z^2)}{6\Gamma(1+\alpha)\Gamma(2-\alpha)T^{1-\alpha}}\Delta^{1-\alpha}.
\end{array}
\end{equation}
Noting that Eq. ( {\color{blue} \ref{eq32}} ) is valid on the condition $\Delta \gg \Delta_c$.

When comes to the case $\Delta \ll \tau_c$, we can put the series expansion of Mittag-Leffler function $E_{\alpha,\beta}[-(m/\tau_c)^\alpha]=\sum_{n=0}^{\infty}\frac{[-(m/\tau_c)^\alpha]^n}{\Gamma(\beta+n\alpha)}$ into use. Removing the high-order items, which can be neglected when $\Delta \ll \tau_c$, and keep $n=0$ item, we get the result:
\begin{equation}\label{eq33}
\begin{array}{l l l l}
\langle \overline{\delta^2}(\Delta) \rangle & \approx & \frac{(L_x^2+L_y^2+L_z^2)}{6\Gamma(1+\alpha)\tau_c{^a}T^{1-\alpha}}\Delta.
\end{array}
\end{equation}
Taking $\tau_c=(\frac{L_x^2+L_y^2+L_z^2}{\pi^2 K_\alpha})^{1/\alpha}$ into consideration, we find that:
\begin{equation}\label{eq34}
\begin{array}{l l l l}
\langle \overline{\delta^2}(\Delta) \rangle & \approx & \frac{\pi^2K_\alpha}{6\Gamma(1+\alpha)T^{1-\alpha}}\Delta.
\end{array}
\end{equation}
 Eq. ( {\color{blue} \ref{eq34}} ) is valid for $\Delta \ll \tau_c$.

\section{Numerical simulation}

\subsection{CTRW simulations}
 We simulate CTRW trajectories of unbiased random walks in three dimensions with the reflecting boundaries $1 \le x \le 31$, $1 \le y \le 21$ and $1 \le z \le 11$. The subdiffusion exponent is $\alpha=0.75$. The total measurement  time is $T=10^8$. The lag time is $\Delta = 100$. The PDFs of waiting time and displacement are defined as $w(t)= \frac{\alpha}{(1+t)^{1+\alpha}}$ and $f(x,y,z)=[\delta(x-1)+\delta(x+1)+\delta(y-1)+\delta(y+1)+\delta(z-1)+\delta(z+1)]/6$ respectively. In Fig.1, the time averaged MSD $\overline{\delta^2}$ of $20$ individual trajectories are denoted by thin dotted lines, which show individual time averaged MSD remains independent random variable. The ensemble average of time averaged MSD is represented by a thick green dotted line, which has a distinct crossover at $\tau_c$ from $ \langle \overline{\delta^2} \rangle \sim\Delta$ in short lag time to $ \langle \overline{\delta^2} \rangle \sim\Delta^{1-\alpha}$ in long lag time, where $\tau_c = (\frac{L_x^2+L_y^2+L_z^2}{\pi^2 K_\alpha})^{1/\alpha} = 10383$, and $ K_\alpha=\frac{\langle \delta r^2 \rangle }{2\Gamma(1-\alpha)}$. The same behaviour was also obtained in previous one-dimensional simulations $^{[{\color{blue} 39,50-53}]}$.

\begin{figure}[!ht]
\centering
\includegraphics[width=15cm,height=10cm]{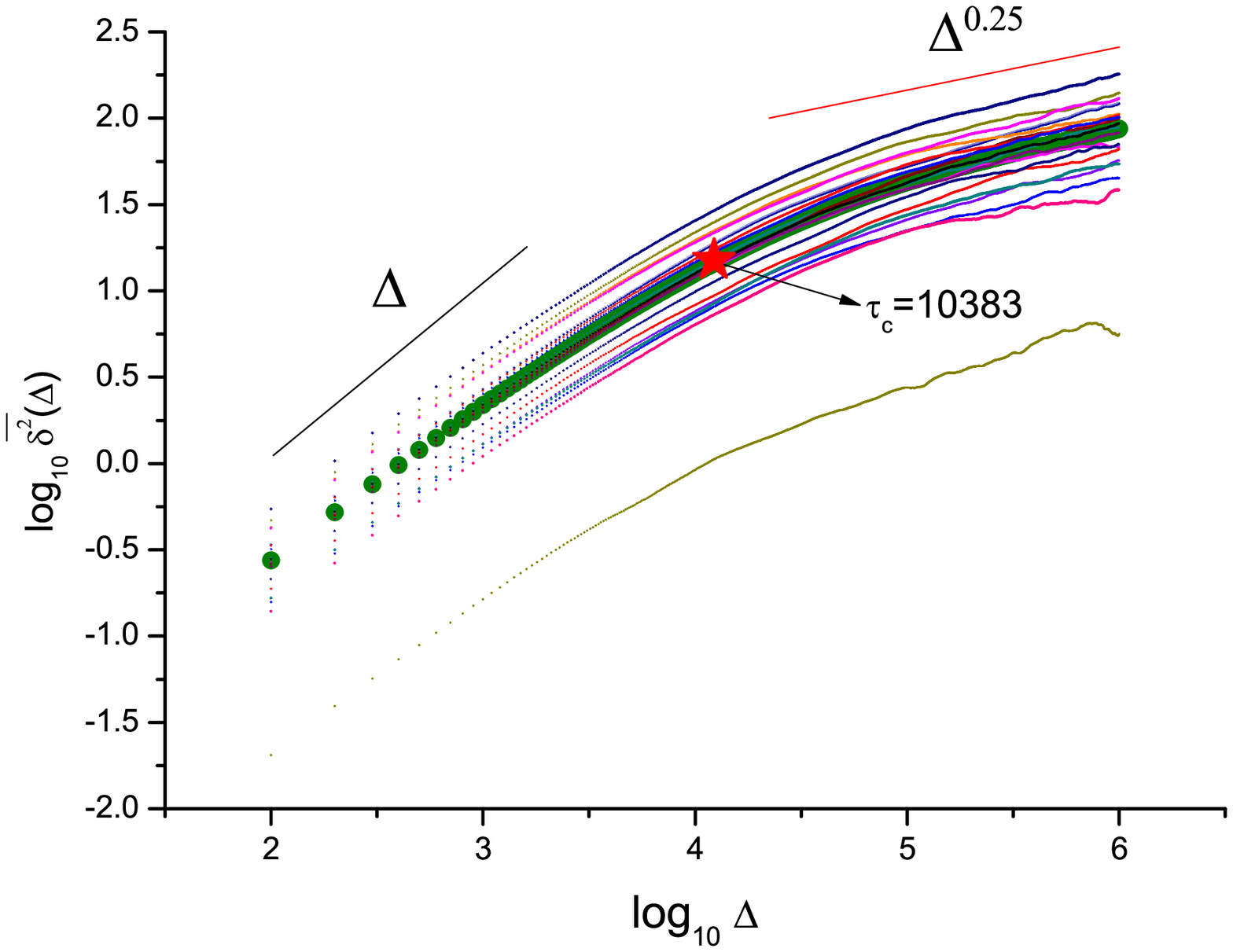}
\begin{center}
\parbox{15.5cm}{\small{\bf Fig.1.}  CTRW simulation of time averaged MSD as a function of $\Delta$ within 3D confined space. The walkers are entrapped in the space $1 \le x \le 31$, $1 \le y \le 21$ and $1 \le z \le 11$, with PDF of waiting time $w(t)= \frac{\alpha}{(1+t)^{1+\alpha}}$ and an unbiased PFD of displacement $f(x,y,z)=[\delta(x-1)+\delta(x+1)+\delta(y-1)+\delta(y+1)+\delta(z-1)+\delta(z+1)]/6$. Here $\Delta =100$, $\alpha=0.75$ and the total measurement time $T=10^8$. All simulations are started at $x=1, y=1, z=1$. A random waiting time $ t $ with the PDF $w(t)=\frac{\alpha}{(1+t)^{1+\alpha}}$ can be generated from $ t=r^{-1/\alpha}-1$, where r is an uniformly distributed random number. Ensemble average of time averaged MSD are taken from 20 trajectories (the thick green dotted line). }
\end{center}
\end{figure}

We then compare the CTRW simulation with the theoretical prediction given by  Eq. ( {\color{blue} \ref{eq31}} ). The parameters are the same as Fig. 1. As shows in Fig.2, the simulated ensemble average of time averaged MSD $\langle \overline{\delta^2} \rangle $ ( green dots ), is identical with the theory ( blue dots ), given by function Eq. ( {\color{blue} \ref{eq31}} ). The previous works dominantly discuss the asymptotic behaviours as $T \gg \Delta \gg \tau_c$ and $ \Delta \ll \tau_c$ $^{[{\color{blue} 39,50}]}$, which well agree with our theoretical results.

\begin{figure}[!ht]
\centering
\includegraphics[width=12cm,height=8cm]{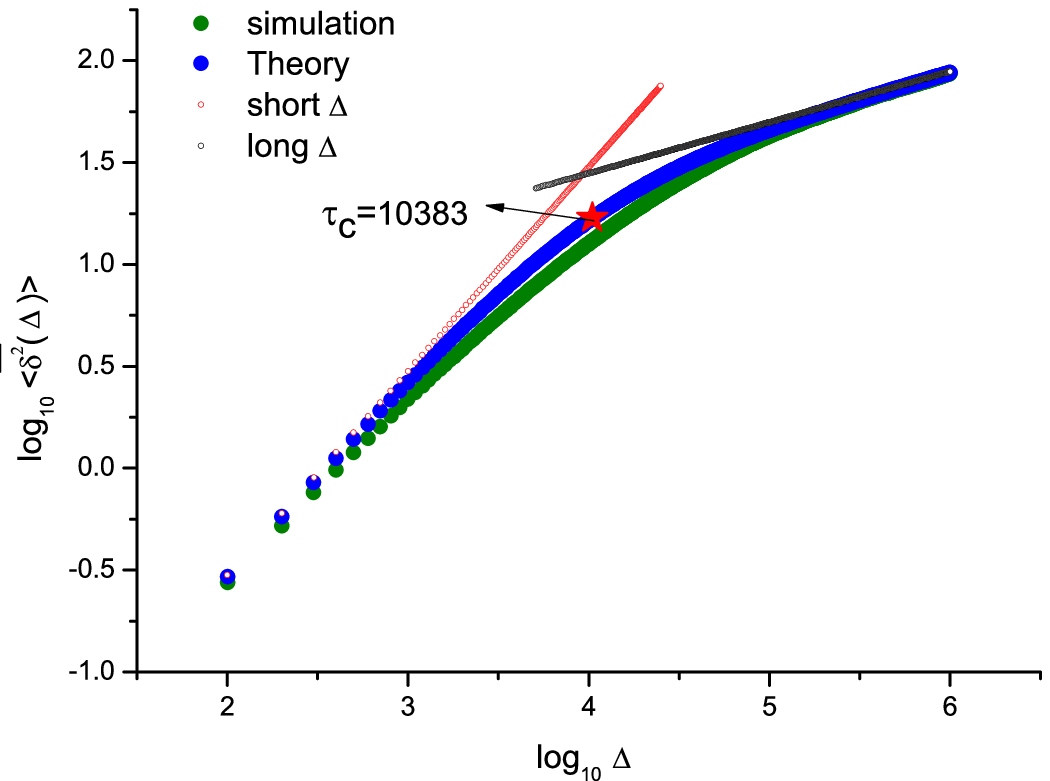}
\begin{center}
\parbox{15.5cm}{\small{\bf Fig.2.}  Comparing CTRW simulation results with the theoretical formula of ensemble average of time averaged MSD $\langle \overline{\delta^2} \rangle $. With $\alpha=0.75$, $\Delta=100$, reflecting boundaries $1 \le x \le 31$, $1 \le y \le 21$ and $1\le z \le 11$ and the total measurement time is $T=10^8$. The green dotted line represents the simulation, and the blue one is theory, as Eq. ( {\color{blue} \ref{eq31}} ). The red line and gray line are the theoretical results in short and long $\Delta$. }
\end{center}
\end{figure}

In contrast, we also do comparison on the behaviors of the ensemble average of time averaged MSD $\langle \overline{\delta^2}(\Delta) \rangle $ for three-dimensional diffusion, with $\langle \overline{\delta^2}_{x}(\Delta) \rangle $, $ \langle \overline{\delta^2}_{y}(\Delta) \rangle $ and $\langle \overline{\delta^2}_{z}(\Delta) \rangle $ for only one direction diffusion of x, y, and z. In Fig. 3, the parameters are also the same as Fig. 1. But note that the crossover points $\tau_c$ are different for every direction, because there are different boundaries in x, y, and z directions.

\begin{figure}[!ht]
\centering
\includegraphics[width=12cm,height=8cm]{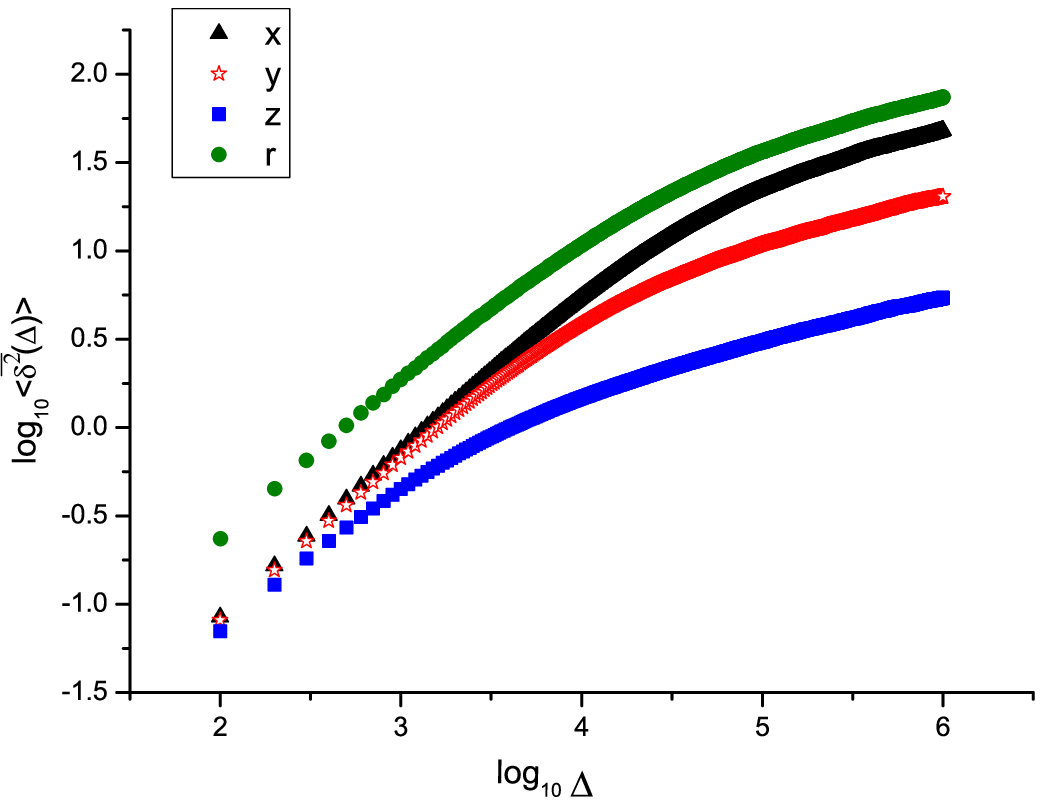}
\begin{center}
\parbox{15.5cm}{\small{\bf Fig.3.}  Simulating the effect of boundaries on ensemble average of time averaged MSD in three different directions. Here, the total measure time $T=10^8$, $\alpha=0.75$, $\Delta=100$ and reflecting boundaries $1 \le x \le 31$, $1 \le y \le 21$ and $1\le z \le 11$. }
\end{center}
\end{figure}

\subsection{Comparison with experiments}
In order to put our theoretical considerations into test, results are presented to compare with the experiment in Fig.4. The details of experiment are reported in the paper ${[{\color{blue} 37}]}$. We treat the cellular inner membrane as reflecting boundaries in our simulation. The boundaries $1 \le x \le 31$, $1 \le y \le 11$ and $1 \le z \le 11$ indicate the relationship that $x$ is almost three times larger than $y$ and $z$, also $y$ roughly equals with $z$, which is observed by the experiment (Fig.1 (b) in paper ${[{\color{blue} 37}]}$ ). Other parameters are as $T=10^7$, $\alpha=0.7$ and $\Delta=2500$. Our results, showing in Fig.4, are obtained by extracting $x$ and $y$ in the three dimensions simulation as the experiment. The simulated results are highly similar to the experiment results.

\begin{figure}[!ht]
\centering
\includegraphics[width=15cm,height=10cm]{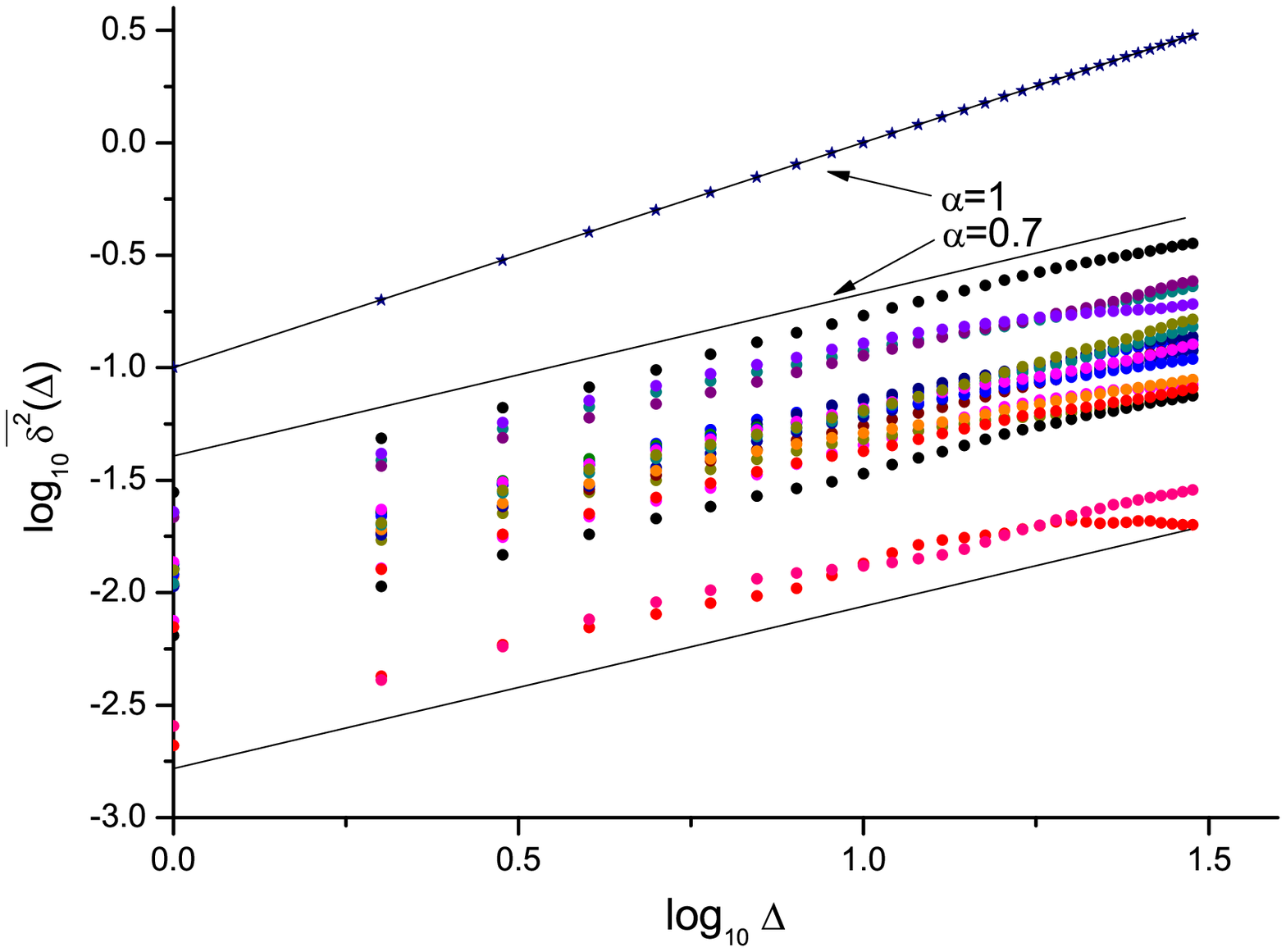}
\begin{center}
\parbox{15.5cm}{\small{\bf Fig.4.} The time averaged MSD $\overline{\delta^2}$ as a function of $\Delta$. Here the total measure time $T=10^7$, $\alpha=0.7$, $\Delta=2500$ and reflecting boundaries $1 \le x \le 31$, $1\le y \le 21$ and $1 \le z \le 11$. }
\end{center}
\end{figure}

As an additional way to characterize subdiffusion and testity the CTRW, we use the data of $x(t)$ trajectories  extracting from simulation of a single particle trajectory in three dimensions to measure the power spectrum$^{[{\color{blue} 54}]}$, with the total measure time $T=10^7$, also $\alpha=0.7$, $\Delta=2500$ and reflecting boundaries $1 \le x \le 31$, $1\le y \le 11$ and $1 \le z \le 11$. The power spectrum of a particle is $P(f)=\left|{X^2(f)}\right|$, which should obey $P(f)\sim f^{-(1+\alpha)}$ when $X(f)$ is the Fourier Transform of particle position x(t). In the same way, we also can calculate the power spectrum for the position y(t) or z(t). The simulation results from CTRW are in a highly agreement with experimental data$^{[{\color{blue} 37}]}$, as showing in Fig.5.

\begin{figure}[!ht]
\centering
\includegraphics[width=15cm,height=10cm]{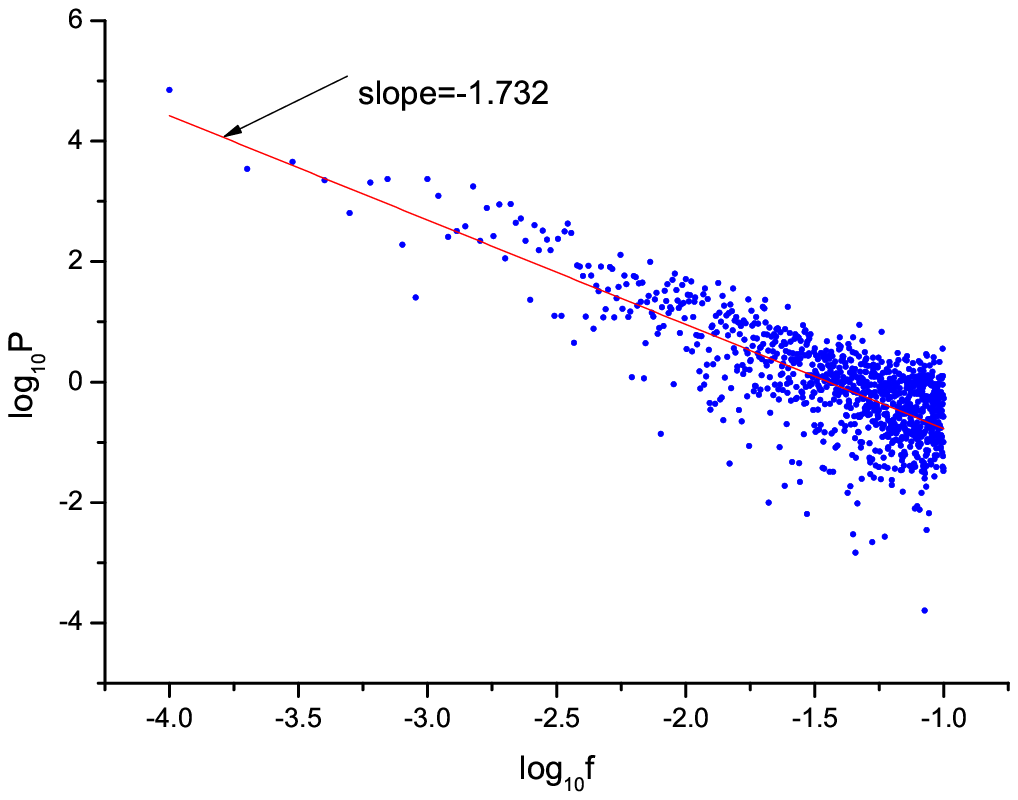}
\begin{center}
\parbox{15.5cm}{\small{\bf Fig.5.} Power Spectrum of a single particle position $x(t)$ (blue dots). Using the Welch method gives the results of fitting slope about equal to $-1.732$. The total measure time is $T=10^7$, also $\alpha=0.7$, $\Delta=2500$ and reflecting boundaries $1 \le x \le 31$, $1\le y \le 11$ and $1 \le z \le 11$. }
\end{center}
\end{figure}

\section{Conclusion}
We investigate the time averaged mean square displacement ( MSD ) of 3D single particle trajectory using the continuous time random walk model (CTRW). Theoretical analysis and numerical simulation show that the CTRW model is suitable for description of the subdiffusion of mRNA in live E. coli under the confined and crowded environment. Theoretical expression of ensemble average of the time averaged MSD is in agreement with simulated results. The simulated results are also well consistent with experimental data. The 3D trajectory and analysis can help to determinate more exactly the microscopic mechanism of real single biomolecules diffusion in living cells.

%%%% �ο������Ű���ʽ��

%
\end{CJK*}  %% �������ġ����ġ�����ʹ�û���
\end{document}